\documentclass[12pt]{article}%
\usepackage{latexsym}
\usepackage{amsmath}
\usepackage{graphicx}
\usepackage{amsfonts}
\usepackage{amssymb}%
\providecommand{\U}[1]{\protect\rule{.1in}{.1in}}
\newcommand{\tri}[7]{\put(#1,#2){\makebox(#3,#4)[t]{$#5$}}
\put(#1,#2){\line(1,1){#6}}
\put(#7,#2){\line(-1,1){#6}}}

\def\rahmen#1{{\dimen0=#1 \advance\dimen0 by -0.8pt\relax
\mathop{\mkern0.5\thinmuskip
\vbox{\hrule width #1
\hbox to #1{\vrule
\hfill
\vrule height\dimen0 width 0pt
\vrule}\hrule width #1}\mkern0.5\thinmuskip}}}

\newtheorem{theorem}{Theorem}
\newtheorem{corollary}[theorem]{Corollary}
\newtheorem{lemma}[theorem]{Lemma}
\begin{document}

\title{Recursive decoding of Reed-Muller codes \thanks{This research is supported by
NSF grant No. NCR-9703844.}}
\author{{\normalsize Ilya Dumer}\\{\small Dept. of Electrical Engineering }\\{\small University of California}\\{\small Riverside, CA 92521}\\{\small dumer@ee.ucr.edu}}
\date{}
\maketitle

\begin{abstract}
New soft- and hard decision decoding algorithms are presented for general
Reed-Muller codes $\left\{\genfrac{}{}{0pt}{}{m}{r}\right\}  $ 
of length $2^{m}$ and distance $2^{m-r}$. We use Plotkin $(u,u+v)$
construction and decompose code $\left\{\genfrac{}{}{0pt}{}{m}{r}\right\}  $ 
onto subblocks $u\in\left\{\genfrac{}{}{0pt}{}{m-1}{r}\right\}  $ 
and $v\in\left\{\genfrac{}{}{0pt}{}{m-1}{r-1}\right\}  .$ 
In decoding, we first try to find a subblock $v$ from the better
protected code and then proceed with the block $u$. The likelihoods of the
received symbols are recalculated in a way similar to belief propagation.
Thus, decoding is relegated to the two constituent codes. We repeat this
recursion and execute decoding only at the end nodes $\left\{\genfrac{}{}{0pt}{}{j}{1}\right\}  $ 
and $\left\{\genfrac{}{}{0pt}{}{j}{j-1}\right\}  $. 
The overall complexity has low order of $n\log n.$ It is shown
that this decoding substantially outperforms other algorithms of polynomial
complexity known for RM codes. 
In particular, for medium and high code rates, the algorithm corrects most error patterns of weight $d\ln d/2.$

\end{abstract}

\thispagestyle{empty}

\section{Introduction}

\bigskip\thispagestyle{empty}The Reed-Muller code $RM(r,m)$ \cite{mac}, which
will be denoted below $\left\{
%TCIMACRO{\QATOP{\QTR{bf}{m}}{\QTR{bf}{r}}}%
%BeginExpansion
\genfrac{}{}{0pt}{}{\mathbf{m}}{\mathbf{r}}%
%EndExpansion
\right\}  ,$ has parameters%

\[
n=2^{m},\text{ }k=\sum_{i=0}^{r}{\binom{m}{i}},\text{ }d=2^{m-r}.
\]
To construct this code, consider all polynomials $z\mathbf{(}x_{1},...,x_{m})$
\ of degree $r$ or less taken over $m$ Boolean variables. Then a codeword
$\mathbf{z}$ is an ordered set of $\ $all $2^{m}$ \ values that polynomial
$\ z$ takes on \ these variables. It is also well known \cite{mac} that RM
codes can be designed by repetitive employment of the Plotkin $(\mathbf{u,u+v}%
)$ construction. Here the original block $(\mathbf{u,u+v})\in\left\{
%TCIMACRO{\QATOP{\QTR{bf}{m}}{\QTR{bf}{r}}}%
%BeginExpansion
\genfrac{}{}{0pt}{}{\mathbf{m}}{\mathbf{r}}%
%EndExpansion
\right\}  $ is represented by two subblocks\textbf{\ }$\mathbf{u}\in\left\{
%TCIMACRO{\QATOP{\QTR{bf}{m-1}}{\QTR{bf}{r}}}%
%BeginExpansion
\genfrac{}{}{0pt}{}{\mathbf{m-1}}{\mathbf{r}}%
%EndExpansion
\right\}  $ and $\mathbf{v}\in\left\{
%TCIMACRO{\QATOP{\QTR{bf}{m-1}}{\QTR{bf}{r-1}}}%
%BeginExpansion
\genfrac{}{}{0pt}{}{\mathbf{m-1}}{\mathbf{r-1}}%
%EndExpansion
\right\}  .$ Now we can specify general $(\mathbf{u,u+v})$ construction as
\begin{equation}
\left\{
%TCIMACRO{\QATOP{m}{r}}%
%BeginExpansion
\genfrac{}{}{0pt}{}{m}{r}%
%EndExpansion
\right\}  =\left\{
%TCIMACRO{\QATOP{m-1}{r}}%
%BeginExpansion
\genfrac{}{}{0pt}{}{m-1}{r}%
%EndExpansion
\right\}  ,\left\{
%TCIMACRO{\QATOP{m-1}{r}}%
%BeginExpansion
\genfrac{}{}{0pt}{}{m-1}{r}%
%EndExpansion
\right\}  +\left\{
%TCIMACRO{\QATOP{m-1}{r-1}}%
%BeginExpansion
\genfrac{}{}{0pt}{}{m-1}{r-1}%
%EndExpansion
\right\}  , \label{recur}%
\end{equation}
where the same codeword $\mathbf{u}\in\left\{
%TCIMACRO{\QATOP{\QTR{bf}{m-1}}{\QTR{bf}{r}}}%
%BeginExpansion
\genfrac{}{}{0pt}{}{\mathbf{m-1}}{\mathbf{r}}%
%EndExpansion
\right\}  $ taken on both halves. In turn, we can split $\mathbf{u}$ and
$\mathbf{v}$ further and obtain successively all codes $\left\{
%TCIMACRO{\QATOP{\QTR{bf}{m}}{\QTR{bf}{r}}}%
%BeginExpansion
\genfrac{}{}{0pt}{}{\mathbf{m}}{\mathbf{r}}%
%EndExpansion
\right\}  $. On step $m,$ we can take the repetition code $\left\{
%TCIMACRO{\QATOP{\QTR{bf}{m}}{\QTR{bf}{0}}}%
%BeginExpansion
\genfrac{}{}{0pt}{}{\mathbf{m}}{\mathbf{0}}%
%EndExpansion
\right\}  $ and full code $\left\{
%TCIMACRO{\QATOP{\QTR{bf}{m}}{\QTR{bf}{m}}}%
%BeginExpansion
\genfrac{}{}{0pt}{}{\mathbf{m}}{\mathbf{m}}%
%EndExpansion
\right\}  ,$ while all other codes $\left\{
%TCIMACRO{\QATOP{\QTR{bf}{m}}{\QTR{bf}{r}}}%
%BeginExpansion
\genfrac{}{}{0pt}{}{\mathbf{m}}{\mathbf{r}}%
%EndExpansion
\right\}  $ are obtained by recursion (\ref{recur}) from the previous step.
Thus, any code $\left\{
%TCIMACRO{\QATOP{\QTR{bf}{m}}{\QTR{bf}{r}}}%
%BeginExpansion
\genfrac{}{}{0pt}{}{\mathbf{m}}{\mathbf{r}}%
%EndExpansion
\right\}  $ can be mapped onto the $(m,r)$-node of the Pascal triangle. Given
any intermediate node $(j,s),$ we move up and left to achieve the node
$(j-1,s-1);$ or up and right to reach the node $(j-1,s).$ Note also that we
can end our recursion on any level. In the sequel, we terminate our splitting
at the nodes $(j,1)$ (corresponding to biortogonal codes ) or at the nodes
$(j,j-1)$ (corresponding to the single parity check code). This is
schematically shown in Fig. 1 for RM codes of the seventh order.\smallskip
\begin{figure}[ptbh]
\begin{picture}(300,120)(-25,0)
\tri{138}{100}{24}{20}{2,1}{12}{162}
\tri{126}{78}{24}{20}{3,1}{12}{150}
\tri{150}{78}{24}{20}{3,2}{12}{174}
\tri{114}{56}{24}{20}{4,1}{12}{138}
\tri{138}{56}{24}{20}{4,2}{12}{162}
\tri{162}{56}{24}{20}{4,3}{12}{186}
\tri{102}{34}{24}{20}{5,1}{12}{126}
\tri{126}{34}{24}{20}{5,2}{12}{150}
\tri{150}{34}{24}{20}{5,3}{12}{174}
\tri{174}{34}{24}{20}{5,4}{12}{198}
\tri{90}{12}{24}{20}{6,1}{12}{114}
\tri{114}{12}{24}{20}{6,2}{12}{138}
\tri{138}{12}{24}{20}{6,3}{12}{162}
\tri{162}{12}{24}{20}{6,4}{12}{186}
\tri{186}{12}{24}{20}{6,5}{12}{210}
\tri{78}{-10}{24}{20}{7,1}{0}{102}
\tri{102}{-10}{24}{20}{7,2}{0}{126}
\tri{126}{-10}{24}{20}{7,3}{0}{150}
\tri{150}{-10}{24}{20}{7,4}{0}{174}
\tri{174}{-10}{24}{20}{7,5}{0}{198}
\tri{198}{-10}{24}{20}{7,6}{0}{222}
\end{picture}
\caption{RM codes of length 128 on Pascal Triangle}%
\label{fig:pas7}%
\end{figure}
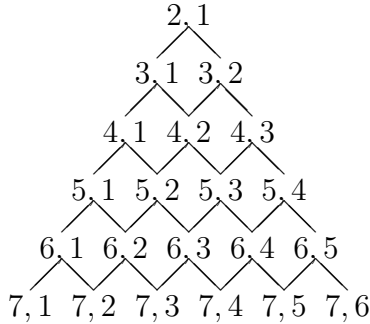

Now let $I%
%TCIMACRO{\QATOP{m}{r}}%
%BeginExpansion
\genfrac{}{}{0pt}{}{m}{r}%
%EndExpansion
$ denote the block of $\ k$ information bits \ used to encode a vector
$(\mathbf{u},\mathbf{u}+\mathbf{v})\in\left\{
%TCIMACRO{\QATOP{\QTR{bf}{m}}{\QTR{bf}{r}}}%
%BeginExpansion
\genfrac{}{}{0pt}{}{\mathbf{m}}{\mathbf{r}}%
%EndExpansion
\right\}  .$ It is also important that recursion (\ref{recur}) splits $I%
%TCIMACRO{\QATOP{m}{r}}%
%BeginExpansion
\genfrac{}{}{0pt}{}{m}{r}%
%EndExpansion
$ into two subblocks $I%
%TCIMACRO{\QATOP{m-1}{r}}%
%BeginExpansion
\genfrac{}{}{0pt}{}{m-1}{r}%
%EndExpansion
$ and $I%
%TCIMACRO{\QATOP{m-1}{r-1}}%
%BeginExpansion
\genfrac{}{}{0pt}{}{m-1}{r-1}%
%EndExpansion
$ that correspond to vectors $\mathbf{u}$ and $\mathbf{v,}$ respectively. In
this way, the new information strings are split again until we arrive at the
end nodes $(j,1)$ or $(j,j-1).$ Thus, any specific codeword can be encoded
from (multiple) information strings assigned to the end nodes.

\section{Background}

Despite relatively bad code distance, RM codes have been considered in
numerous publications thanks to efficient decoding procedures.
\textit{Majority decoding} developed in \cite{ree} (see also \cite{gor},
\cite{kol}, \cite{mas}, and \cite{rud}) has complexity order at most $nk$ and
corrects all error patterns of weight below $d/2$. It is also known \cite{kri}
that majority decoding corrects many error patterns of higher weights and can
be further improved for soft-decision channels \cite{dum3}.

To decrease decoding complexity, \textit{recursive algorithms} were also
developed on both hard- (\cite{lit}) and soft-decision (\cite{kab}) channels.
The algorithms provide for bounded distance decoding and have the lowest
complexity order $n\min(r,m-r)$ known for RM codes to date. \thinspace
\thinspace Simulation results presented in \cite{bos} also showed that
recursive soft-decision algorithms can increase decoding domain of bounded
distance decoding. Finally, efficient \textit{permutation \ }algorithm
considered in \cite{sid} for codes $\left\{
%TCIMACRO{\QATOP{\QTR{bf}{m}}{\QTR{bf}{2}}}%
%BeginExpansion
\genfrac{}{}{0pt}{}{\mathbf{m}}{\mathbf{2}}%
%EndExpansion
\right\}  ,$ gives a slightly higher complexity $O(n^{2}m)$ while correcting
most error patterns of a higher weight $n(1-h)/2,$ where $h$ has a vanishing
order of $(m/n)^{1/4}$ as $m\rightarrow\infty.$ \

In this paper, we wish to further develop recursive algorithms for RM codes
and improve their performance both on short and long lengths. Such an
improvement is especially important for short and moderate lengths on which RM
codes are on par with the best codes known to date. Our goal is to use the
same recursive presentation on Pascal triangle that is used \ above in code
design. Namely, given an output $\mathbf{y=(y}^{\prime}\mathbf{,y}%
^{\prime\prime}\mathbf{)}$ with halves $\mathbf{y}^{\prime}$ and
$\mathbf{y}^{\prime\prime},$ we wish to perform two steps:

$1.$ \textit{combine }$\mathbf{y}^{\prime}$\textbf{\ }and\textbf{\ }%
$\mathbf{y}^{\prime\prime}$ to find $\mathbf{v}\in\left\{
%TCIMACRO{\QATOP{\QTR{bf}{m-1}}{\QTR{bf}{r-1}}}%
%BeginExpansion
\genfrac{}{}{0pt}{}{\mathbf{m-1}}{\mathbf{r-1}}%
%EndExpansion
\right\}  .$

$2.$ \textit{combine } $\mathbf{(y}^{\prime}\mathbf{,y}^{\prime\prime
}\mathbf{)}$ and $\mathbf{(0,v)}$ to find $\mathbf{u}\in\left\{
%TCIMACRO{\QATOP{\QTR{bf}{m-1}}{\QTR{bf}{r}}}%
%BeginExpansion
\genfrac{}{}{0pt}{}{\mathbf{m-1}}{\mathbf{r}}%
%EndExpansion
\right\}  .$

\noindent In turn, shorter codes $\left\{
%TCIMACRO{\QATOP{\QTR{bf}{m-1}}{\QTR{bf}{r-1}}}%
%BeginExpansion
\genfrac{}{}{0pt}{}{\mathbf{m-1}}{\mathbf{r-1}}%
%EndExpansion
\right\}  $ and $\left\{
%TCIMACRO{\QATOP{\QTR{bf}{m-1}}{\QTR{bf}{r}}}%
%BeginExpansion
\genfrac{}{}{0pt}{}{\mathbf{m-1}}{\mathbf{r}}%
%EndExpansion
\right\}  $ will be split further, while actual decoding procedures will be
relegated to the end nodes. In the following sections, this procedure is
discussed in more detail.

\section{Decoding}

Consider now the channel with Gaussian noise $\mathcal{N}(0,\sigma^{2})$ and
probability density function
\begin{equation}
G(y)=(1/\sqrt{2\pi}\sigma)e^{-y^{2}/2\sigma^{2}}. \label{w3}%
\end{equation}
The two symbols $0$ and $1$ are transmitted as $+1$ and $-1.$ These two take
arbitrary real values $y$ at the receiver end with probability densities
$G(y+1)$ and $G(y-1),$ respectively. In hard decision reception, we arrive at
the BSC with transition error probability $p=Q(1/\sigma),$ where
\[
Q(x)=\int_{x}^{\infty}e^{-y^{2}/2}dy/\sqrt{2\pi}.
\]
In brief, we call these two channels AWGN$(\sigma^{2})$ and BSC$(p)$ respectively.

Below we use the code $\left\{
%TCIMACRO{\QATOP{\QTR{bf}{m}}{\QTR{bf}{r}}}%
%BeginExpansion
\genfrac{}{}{0pt}{}{\mathbf{m}}{\mathbf{r}}%
%EndExpansion
\right\}  $ of length $n=2^{m\text{ }}.$ Suppose that the codeword
$\mathbf{z}$ is transmitted and\textbf{\ }$\mathbf{y}\in\mathcal{R}^{n}$ is
received. Given any output signal $y\in\mathcal{R},$ we can find the
posterior\textit{\ }probabilities $p\overset{def}{=}p(1|y)$ and $q\overset
{def}{=}p(0|y)$. By using the Bayes' rule we find
\begin{equation}
p=e^{-g/2}/(e^{g/2}+e^{-g/2}),\;q=e^{g/2}/(e^{g/2}+e^{-g/2}). \label{w5}%
\end{equation}
Here $g$ is the \textit{likelihood }of symbol\textit{\ }$0:$%
\begin{equation}
g=\log(q/p)=2y/\sigma^{2}. \label{reli}%
\end{equation}
Finally, we introduce the \textit{spread } $h$ (which is the hyperbolic
tangent of $g)$ between the two probabilities $q$ and $p:$%
\begin{equation}
h=q-p=(e^{g/2}-e^{-g/2})/(e^{g/2}+e^{-g/2})=\tanh(g). \label{spr}%
\end{equation}
Given an output vector $\mathbf{y=(}y_{1},...,y_{n}\mathbf{),}$ we can find
the quantities $q_{j},$ $p_{j},$ $h_{j},$ and $g_{j}$ for any position $j.$ In
decoding, we will use the original vector $\mathbf{y},$ as well as the
corresponding vectors $\mathbf{h}=(h_{1},...,h_{n})$ and $\mathbf{g=}%
(g_{1},...,g_{n})$. Note that $\tanh(g)$ is a one-to-one mapping. Therefore
the three vectors are interchangeable:
\begin{equation}
\mathbf{y\Leftrightarrow g\Leftrightarrow h.} \label{equi}%
\end{equation}
We denote such a decoding $\mathbf{z=}\Psi_{r}^{m}(\mathbf{y})=\Psi_{r}%
^{m}(\mathbf{h})=\Psi_{r}^{m}(\mathbf{g}),$ where $\mathbf{z}\in\left\{
%TCIMACRO{\QATOP{\QTR{bf}{m}}{\QTR{bf}{r}}}%
%BeginExpansion
\genfrac{}{}{0pt}{}{\mathbf{m}}{\mathbf{r}}%
%EndExpansion
\right\}  $ is our decoding result.

In ML decoding, we can first consider the string of hard decision outputs%

\begin{equation}
a_{j}=\left\{
\begin{array}
[c]{ll}%
0, & \text{if}\;y_{j}\geq0,\\
1, & \text{if}\;y_{j}<0,
\end{array}
\right.  \label{s4}%
\end{equation}
and try to find the most reliable codeword
\[
\mathbf{z}^{\ast}\mathbf{:}\sum_{j:z_{j}^{\ast}\neq a_{j}}| g_{j}|\leq
\sum_{j:z_{j}\neq a_{j}}| g_{j}|
\]
among all codewords $\mathbf{z}\in\left\{
%TCIMACRO{\QATOP{\QTR{bf}{m}}{\QTR{bf}{r}}}%
%BeginExpansion
\genfrac{}{}{0pt}{}{\mathbf{m}}{\mathbf{r}}%
%EndExpansion
\right\}  .$ In our decoding below we also wish to minimize $\sum_{j:z_{j}\neq
a_{j}}| g_{j}|.$ However, this will be done on the premise that $\mathbf{y}$
is not heavily corrupted by noise. The corresponding threshold levels will be
defined for BSC$(p)$ and AWGN$(\sigma^{2})$ in Theorems 4 and 5, respectively.

\subsection{Recalculating the probabilities}

Let $\mathbf{z}^{\prime}$ and $\mathbf{z}^{\prime\prime}$ denote the left- and
right halves of any vector $\mathbf{z}\in\left\{
%TCIMACRO{\QATOP{\QTR{bf}{m}}{\QTR{bf}{r}}}%
%BeginExpansion
\genfrac{}{}{0pt}{}{\mathbf{m}}{\mathbf{r}}%
%EndExpansion
\right\}  $. We also use odd positions $j=2s-1$ on the left half and their
even counterparts $2s$ on the right one for any $s=1,...,n/2.$ Similar
notations are used for all other vectors, say $\mathbf{y,h}$ and $\mathbf{g.}$
Given any vector $\mathbf{z=(u,u+v)}$ at the transmitter end, we can find
\[
\mathbf{v=z}^{\prime}+\mathbf{z}^{\prime\prime}(\operatorname{mod}%
2)\mathbf{=(}z_{1}+z_{2},z_{3}+z_{4},...,z_{n-1}+z_{n}).
\]
By contrast, at the receiver end we know only the strings $\mathbf{p=(}%
p_{1},...,p_{n}),$ $\mathbf{g,}$ and $\mathbf{h}$ that define the probability
distribution on the transmitted symbols$\mathbf{.}$ Our first problem is to
find the spread $\mathbf{h}_{\ast}$ on vectors $\mathbf{v=z}^{\prime
}+\mathbf{z}^{\prime\prime}$ given the original spread $\mathbf{h}=$
$(\mathbf{h}^{\prime},\mathbf{h}^{\prime\prime})$ on vectors $\mathbf{z=(z}%
^{\prime},\mathbf{z}^{\prime\prime}).$

\begin{lemma}
(addition $\operatorname{mod}2).$ Vectors $\{\mathbf{z}^{\prime}%
\mathbf{+z}^{\prime\prime}\}$ $(\operatorname{mod}2)$ have the spread
\[
\mathbf{h}_{\ast}=\mathbf{h}^{\prime}\mathbf{h }^{\prime\prime}=( h_{1}
h_{2},..., h _{n-1} h_{n}).
\]
\noindent
\end{lemma}

\noindent\noindent Given an output $\mathbf{y},$ we now find the probability
spreads $\mathbf{h}$ and $\mathbf{h}^{\ast}$ on vectors $\mathbf{z}$ and
$\mathbf{z}^{\prime}\mathbf{+z}^{\prime\prime}.$ Then our decoding $\Psi
_{r}^{m}(\mathbf{h})\mathbf{=(u,u+v)}$ can first try to find $\mathbf{v}%
=\Psi_{r-1}^{m-1}(\mathbf{h}^{\ast})$. Once vector $\mathbf{v}$ is found, we
wish to find the remaining block $\mathbf{(u,u+v)}+\mathbf{(0,v)=(u,u)}$ in
the second step of our decoding. \ Here we need to replace original symbols
$z_{2s}$ by symbols $z_{2s}+v_{s}.$ Correspondingly, the latter have
likelihoods
\[
\overline{g}_{2s}^{_{{}}}=\left\{
\begin{array}
[c]{ll}%
g_{2s}^{_{{}}}, & \text{if}\;v_{s}=0,\\
-g_{2s}^{_{{}}}, & \text{if}\;v_{s}=1.
\end{array}
\right.
\]
In other words, we change the sign of $g_{2s}^{_{{}}}$ whenever $v_{2s}=1.$
The result is the string $\overline{\mathbf{g}}(\mathbf{u,u})=(\overline
{\mathbf{g}}^{\prime},\overline{\mathbf{g}}^{\prime\prime}),$ where the left
half $\overline{\mathbf{g}}^{\prime}$ is taken from the original vector
$\mathbf{g}$ and equals $\mathbf{g}^{\prime}\mathbf{.}$ The string
$\overline{\mathbf{g}}(\mathbf{u,u})$ represents likelihoods for arbitrary
vectors $\mathbf{(u,u)}$ with two equal parts. Then we find the
likelihoods\textit{\ } $\noindent\mathbf{g}_{\ast}(\mathbf{u})$ given the two
estimates $\overline{\mathbf{g}}^{\prime},\overline{\mathbf{g}}^{\prime\prime
}.$

\begin{lemma}
(repetition). A string of \ likelihoods $(\overline{\mathbf{g}}^{\prime
},\overline{\mathbf{g}}^{\prime\prime})$ defined on repeated vectors
$(\mathbf{u,u})$ gives for vectors $\mathbf{u}$ the string of likelihoods%

\[
\noindent\mathbf{g}_{\ast}(\mathbf{u})=\overline{\mathbf{g}}^{\prime
}\mathbf{+}\overline{\mathbf{g}}^{\prime\prime}.\smallskip
\]

\end{lemma}

\bigskip Once the likelihoods $\mathbf{g}_{\ast}$ are found, we execute
decoding $\Psi%
%TCIMACRO{\QATOP{\QTR{bf}{m-1}}{\QTR{bf}{r}}}%
%BeginExpansion
\genfrac{}{}{0pt}{}{\mathbf{m-1}}{\mathbf{r}}%
%EndExpansion
(\mathbf{g}_{\ast})$ that finds $\mathbf{u\in}\left\{
%TCIMACRO{\QATOP{\QTR{bf}{m-1}}{\QTR{bf}{r}}}%
%BeginExpansion
\genfrac{}{}{0pt}{}{\mathbf{m-1}}{\mathbf{r}}%
%EndExpansion
\right\}  $. We perform both decodings $\Psi%
%TCIMACRO{\QATOP{\QTR{bf}{m-1}}{\QTR{bf}{r-1}}}%
%BeginExpansion
\genfrac{}{}{0pt}{}{\mathbf{m-1}}{\mathbf{r-1}}%
%EndExpansion
(\mathbf{h}_{\ast})$ and $\Psi%
%TCIMACRO{\QATOP{\QTR{bf}{m-1}}{\QTR{bf}{r}}}%
%BeginExpansion
\genfrac{}{}{0pt}{}{\mathbf{m-1}}{\mathbf{r}}%
%EndExpansion
(\mathbf{g}_{\ast})$ in a recursive way. In this process, we only recalculate
spreads $\mathbf{h} $ and likelihoods $\mathbf{g}$ \ while using new (shorter)
codes. Our recursion moves along the edges of Pascal triangle until
recalculated vectors $\mathbf{h}$ and $\mathbf{g}$ arrive at the end nodes
$(j,1)$ and $(j,j-1).$ At the end nodes, we perform ML decoding. Now we can
describe algorithm $\Psi%
%TCIMACRO{\QATOP{\QTR{bf}{m}}{\QTR{bf}{r}}}%
%BeginExpansion
\genfrac{}{}{0pt}{}{\mathbf{m}}{\mathbf{r}}%
%EndExpansion
$ in a general soft-decision setting.

\subsection{Recursive decoding for codes $\left\{
%TCIMACRO{\QATOP{\QTR{bf}{m}}{\QTR{bf}{r}}}%
%BeginExpansion
\genfrac{}{}{0pt}{}{\mathbf{m}}{\mathbf{r}}%
%EndExpansion
\right\}  $}

\textbf{1}. Receive vector $\mathbf{y}\in\mathcal{R}^{n}.$ Calculate
$\mathbf{g=g}%
%TCIMACRO{\QATOP{m}{r}}%
%BeginExpansion
\genfrac{}{}{0pt}{}{m}{r}%
%EndExpansion
$ and $\mathbf{h=h}%
%TCIMACRO{\QATOP{m}{r}}%
%BeginExpansion
\genfrac{}{}{0pt}{}{m}{r}%
%EndExpansion
\mathbf{\ }$according to (\ref{reli}) and (\ref{spr}). \textit{Call procedure}
$\Psi%
%TCIMACRO{\QATOP{m}{r}}%
%BeginExpansion
\genfrac{}{}{0pt}{}{m}{r}%
%EndExpansion
.$ Output decoded vector\textbf{\ }$\mathbf{z}%
%TCIMACRO{\QATOP{m}{r}}%
%BeginExpansion
\genfrac{}{}{0pt}{}{m}{r}%
%EndExpansion
\in\left\{
%TCIMACRO{\QATOP{\QTR{bf}{m}}{\QTR{bf}{r}}}%
%BeginExpansion
\genfrac{}{}{0pt}{}{\mathbf{m}}{\mathbf{r}}%
%EndExpansion
\right\}  $ and its information set $I%
%TCIMACRO{\QATOP{m}{r}}%
%BeginExpansion
\genfrac{}{}{0pt}{}{m}{r}%
%EndExpansion
$.

\quad

\noindent\textbf{2.} \textit{Procedure} $\Psi_{s}^{j}.$ Input $\mathbf{h}%
_{s}^{j}$\textbf{\ }and\textbf{\ }$\mathbf{g}_{s}^{j}.$

\quad

\noindent\textbf{2.1}. Find $\mathbf{h}_{s-1}^{j-1}=(\mathbf{h}_{s}%
^{j})^{\prime}\cdot(\mathbf{h}_{s}^{j})^{\prime\prime}.$ Go to 3 if $s=2$ or
\textit{call } $\Psi_{s-1}^{j-1}$ otherwise.\smallskip

\quad

\noindent\textbf{2.2.} Find $\mathbf{g}%
%TCIMACRO{\QATOP{j-1}{s}}%
%BeginExpansion
\genfrac{}{}{0pt}{}{j-1}{s}%
%EndExpansion
=(\overline{\mathbf{g}}_{s}^{j})^{\prime}+\mathbf{(}\overline{\mathbf{g}}%
_{s}^{j})^{\prime\prime}\mathbf{.}$ Go to 4 if $s=j-2$ or \textit{call } $\Psi%
%TCIMACRO{\QATOP{j-1}{s}}%
%BeginExpansion
\genfrac{}{}{0pt}{}{j-1}{s}%
%EndExpansion
$otherwise.\smallskip

\quad

\noindent\textbf{2.3.} Find $I%
%TCIMACRO{\QATOP{j}{s}}%
%BeginExpansion
\genfrac{}{}{0pt}{}{j}{s}%
%EndExpansion
=I%
%TCIMACRO{\QATOP{j-1}{s-1}}%
%BeginExpansion
\genfrac{}{}{0pt}{}{j-1}{s-1}%
%EndExpansion
\cup I%
%TCIMACRO{\QATOP{j-1}{s}}%
%BeginExpansion
\genfrac{}{}{0pt}{}{j-1}{s}%
%EndExpansion
$ and vector $\mathbf{z}%
%TCIMACRO{\QATOP{j}{s}}%
%BeginExpansion
\genfrac{}{}{0pt}{}{j}{s}%
%EndExpansion
=(\mathbf{z}%
%TCIMACRO{\QATOP{j-1}{s}}%
%BeginExpansion
\genfrac{}{}{0pt}{}{j-1}{s}%
%EndExpansion
,\mathbf{z}%
%TCIMACRO{\QATOP{j-1}{s}}%
%BeginExpansion
\genfrac{}{}{0pt}{}{j-1}{s}%
%EndExpansion
+$ $\mathbf{z}%
%TCIMACRO{\QATOP{j-1}{s-1}}%
%BeginExpansion
\genfrac{}{}{0pt}{}{j-1}{s-1}%
%EndExpansion
).$ \textit{Return }vector $\mathbf{z}%
%TCIMACRO{\QATOP{j}{s}}%
%BeginExpansion
\genfrac{}{}{0pt}{}{j}{s}%
%EndExpansion
\in\left\{
%TCIMACRO{\QATOP{\QTR{bf}{j}}{\QTR{bf}{s}}}%
%BeginExpansion
\genfrac{}{}{0pt}{}{\mathbf{j}}{\mathbf{s}}%
%EndExpansion
\right\}  $ and its information set $I%
%TCIMACRO{\QATOP{j}{s}}%
%BeginExpansion
\genfrac{}{}{0pt}{}{j}{s}%
%EndExpansion
.$

\quad\medskip

\noindent\textbf{3.} Execute ML decoding $\Psi%
%TCIMACRO{\QATOP{j-1}{1}}%
%BeginExpansion
\genfrac{}{}{0pt}{}{j-1}{1}%
%EndExpansion
.$ \textit{Return }$\mathbf{z}%
%TCIMACRO{\QATOP{j-1}{1}}%
%BeginExpansion
\genfrac{}{}{0pt}{}{j-1}{1}%
%EndExpansion
$ and $I%
%TCIMACRO{\QATOP{j-1}{1}}%
%BeginExpansion
\genfrac{}{}{0pt}{}{j-1}{1}%
%EndExpansion
.$

\quad

\noindent\textbf{4.} Execute ML decoding $\Psi_{j-2}^{j-1}.$ \ \textit{Return
}$\mathbf{z}_{j}^{j}$ and $I_{j}^{j}.$

\medskip

\noindent\textit{Qualitative analysis.} Note that increasing the noise power
$\sigma^{2}$ reduces the means of the spreads $\mathbf{h}$ and likelihoods
$\mathbf{g.}$ In particular, it can be shown (see \cite{dum3}) that for large
noise $\sigma\rightarrow\infty,$ the first two moments $Eh$ and $Eh^{2}$ of
the random variable $h=\tanh(2y/\sigma^{2})$ satisfy the relation
\begin{equation}
Eh\sim Eh^{2}\sim\sigma^{-2}. \label{a8}%
\end{equation}
In decoding process, we replace our original decoding $\Psi%
%TCIMACRO{\QATOP{m}{r}}%
%BeginExpansion
\genfrac{}{}{0pt}{}{m}{r}%
%EndExpansion
$ by $\Psi%
%TCIMACRO{\QATOP{m-1}{r-1}}%
%BeginExpansion
\genfrac{}{}{0pt}{}{m-1}{r-1}%
%EndExpansion
.$ The latter operates in a less reliable setting with the lower spread
$\mathbf{h}%
%TCIMACRO{\QATOP{m-1}{r-1}}%
%BeginExpansion
\genfrac{}{}{0pt}{}{m-1}{r-1}%
%EndExpansion
=(\mathbf{h}%
%TCIMACRO{\QATOP{m}{r}}%
%BeginExpansion
\genfrac{}{}{0pt}{}{m}{r}%
%EndExpansion
)^{\prime}\cdot(\mathbf{h}%
%TCIMACRO{\QATOP{m}{r}}%
%BeginExpansion
\genfrac{}{}{0pt}{}{m}{r}%
%EndExpansion
)^{\prime\prime}$ versus\textbf{\ }the original spread\textbf{\ }$\mathbf{h}%
%TCIMACRO{\QATOP{m}{r}}%
%BeginExpansion
\genfrac{}{}{0pt}{}{m}{r}%
%EndExpansion
$. Then the newly derived means $E(\mathbf{h}%
%TCIMACRO{\QATOP{m-1}{r-1}}%
%BeginExpansion
\genfrac{}{}{0pt}{}{m-1}{r-1}%
%EndExpansion
)$ are the componentwise products of \ the corresponding means $E(\mathbf{h}%
%TCIMACRO{\QATOP{m}{r}}%
%BeginExpansion
\genfrac{}{}{0pt}{}{m}{r}%
%EndExpansion
)^{\prime}$ and $E(\mathbf{h}%
%TCIMACRO{\QATOP{m}{r}}%
%BeginExpansion
\genfrac{}{}{0pt}{}{m}{r}%
%EndExpansion
)^{\prime\prime}$. According to (\ref{a8}), this multiplication is equivalent
to replacing the original noise power $\sigma^{2}$ by the larger power
$\sigma^{4}.$ On the other hand, we also increase the relative distance $d/n$
by using a better protected code $\left\{
%TCIMACRO{\QATOP{\QTR{bf}{m-1}}{\QTR{bf}{r-1}}}%
%BeginExpansion
\genfrac{}{}{0pt}{}{\mathbf{m-1}}{\mathbf{r-1}}%
%EndExpansion
\right\}  $ instead of the original code $\left\{
%TCIMACRO{\QATOP{\QTR{bf}{m}}{\QTR{bf}{r}}}%
%BeginExpansion
\genfrac{}{}{0pt}{}{\mathbf{m}}{\mathbf{r}}%
%EndExpansion
\right\}  .$ Our next decoding $\Psi%
%TCIMACRO{\QATOP{m-1}{r}}%
%BeginExpansion
\genfrac{}{}{0pt}{}{m-1}{r}%
%EndExpansion
$ is performed in a better channel. Here we change the original likelihoods
$\mathbf{g}%
%TCIMACRO{\QATOP{m}{r}}%
%BeginExpansion
\genfrac{}{}{0pt}{}{m}{r}%
%EndExpansion
$ for $\mathbf{g}%
%TCIMACRO{\QATOP{m-1}{r}}%
%BeginExpansion
\genfrac{}{}{0pt}{}{m-1}{r}%
%EndExpansion
=(\overline{\mathbf{g}}%
%TCIMACRO{\QATOP{m}{r}}%
%BeginExpansion
\genfrac{}{}{0pt}{}{m}{r}%
%EndExpansion
)^{\prime}\mathbf{+(}\overline{\mathbf{g}}%
%TCIMACRO{\QATOP{m}{r}}%
%BeginExpansion
\genfrac{}{}{0pt}{}{m}{r}%
%EndExpansion
)^{\prime\prime}$. \ As a result, our average likelihoods are doubled. This is
equivalent to the map $\sigma^{2}\Rightarrow\sigma^{2}/2. $ In other words,
code $\left\{
%TCIMACRO{\QATOP{\QTR{bf}{m-1}}{\QTR{bf}{r}}}%
%BeginExpansion
\genfrac{}{}{0pt}{}{\mathbf{m-1}}{\mathbf{r}}%
%EndExpansion
\right\}  $ operates on a channel whose noise power is reduced two times.
However, we also reduce the relative distance by using a weaker code $\left\{
%
%TCIMACRO{\QATOP{\QTR{bf}{m-1}}{\QTR{bf}{r}}}%
%BeginExpansion
\genfrac{}{}{0pt}{}{\mathbf{m-1}}{\mathbf{r}}%
%EndExpansion
\right\}  $ instead of $\left\{
%TCIMACRO{\QATOP{\QTR{bf}{m}}{\QTR{bf}{r}}}%
%BeginExpansion
\genfrac{}{}{0pt}{}{\mathbf{m}}{\mathbf{r}}%
%EndExpansion
\right\}  .$ In the next section, we present the \textit{quantative} results
of decoding performance.

\section{ Summary of results}

We first consider hard decision reception on the channel BSC$(p)$ with
transition error probability $p=(1-h)/2.$

\begin{theorem}
Recursive decoding of codes $\left\{
%TCIMACRO{\QATOP{\QTR{bf}{m}}{\QTR{bf}{r}}}%
%BeginExpansion
\genfrac{}{}{0pt}{}{\mathbf{m}}{\mathbf{r}}%
%EndExpansion
\right\}  $ used on a BSC$(p),$ gives the output bit error probability
$\alpha\leq Q(\mu),$ where
\begin{equation}
\text{ }\mu=2^{(m-r)/2}h^{2^{r-1}}/\sqrt{1-h^{2^{r}}},\quad p=(1-h)/2.
\label{prel6}%
\end{equation}

\end{theorem}

\noindent We then consider asymptotic capacity of recursive algorithms for
long codes $\left\{
%TCIMACRO{\QATOP{\QTR{bf}{m}}{\QTR{bf}{r}}}%
%BeginExpansion
\genfrac{}{}{0pt}{}{\mathbf{m}}{\mathbf{r}}%
%EndExpansion
\right\}  $ as $m\rightarrow\infty$. We consider separately low-rate codes of
fixed order $r$ and those of fixed rate $R.$ The latter implies that
$r/m\rightarrow0.5.$ Let $c$ be any constant exceeding $\ln2.$

\begin{theorem}
For $m\rightarrow\infty,$ recursive decoding of \ codes $\left\{
%TCIMACRO{\QATOP{\QTR{bf}{m}}{\QTR{bf}{r}}}%
%BeginExpansion
\genfrac{}{}{0pt}{}{\mathbf{m}}{\mathbf{r}}%
%EndExpansion
\right\}  $ corrects most error patterns of weight:%

\begin{equation}
t\leq\left\{
\begin{array}
[c]{ll}%
n(1-(cm/d)^{1/2^{r}})/2, & \text{if }\;r=const,\\
d(\ln d-\ln2m)/2, & \text{if }\;0<R<1.
\end{array}
\right.  \label{har-low}%
\end{equation}
with decoding complexity order $n\log n.$
\end{theorem}

\noindent For fixed code rate $R,$ the latter estimate is almost twice the
bound $d\ln d/4$ known \cite{kri} for majority decoding. For low-rate codes of
fixed order $r$ we correct almost $n/2$ errors. In this case, our residual
term $h=1-2t/n$ has vanishing order of $(cm/d)^{1/2^{r}}.$ The latter
substantially reduces the former term $(m/d)^{1/2^{r+1}}$ known \cite{kri} for
majority decoding. Such a performance was known before only for $r=2$ and was
obtained in permutation decoding presented in \cite{sid}. We note that this
threshold $h$ is now obtained for all orders $r$ and is achieved with a lower
complexity order of $n\log n$.

Our next issue is to improve recursive decoding by using soft decision
likelihoods $\mathbf{g.}$ Given any output bit error rate $\alpha<1/2,$ we
compare the corresponding noise powers $\sigma_{s}^{2}$ and $\sigma_{h}^{2} $
\ and transition error probabilities $p_{s}$ and $p_{h}$ that sustain this
probability in soft and hard decision decoding.

\begin{theorem}
Given any output bit error probability $\alpha,$ soft decision recursive
decoding of long codes $\left\{
%TCIMACRO{\QATOP{\QTR{bf}{m}}{\QTR{bf}{r}}}%
%BeginExpansion
\genfrac{}{}{0pt}{}{\mathbf{m}}{\mathbf{r}}%
%EndExpansion
\right\}  $ of fixed order $r$ increases $\pi/2$ times $\alpha$-sustainable
noise power of hard decision decoding:
\begin{equation}
\sigma_{s}^{2}/\sigma_{h}^{2}\rightarrow\pi/2,\quad m\rightarrow\infty.
\label{hsof-low}%
\end{equation}
Soft decision decoding of long codes $\left\{
%TCIMACRO{\QATOP{\QTR{bf}{m}}{\QTR{bf}{r}}}%
%BeginExpansion
\genfrac{}{}{0pt}{}{\mathbf{m}}{\mathbf{r}}%
%EndExpansion
\right\}  $ of fixed code rate $R$ increases $4/\pi$ times $\alpha
$-sustainable transition error probability of hard decision decoding:
\begin{equation}
p_{s}/p_{h}\rightarrow4/\pi,\quad m\rightarrow\infty. \label{hsof-med}%
\end{equation}

\end{theorem}

\smallskip\noindent The above theorem shows that for long low-rate $RM$ codes
we can gain $10\log_{10}(\pi/2)\approx2.0$ dB over hard decision decoding for
any output error rate $\alpha.$ The following corollary concerns the Euclidean
weights of error patterns correctable by our algorithm. We show that for fixed
$r$ we exceed the bounded distance decoding weight $\sqrt{d}$ more than
$2^{r/2}$ times. For fixed code rate $R,$ we have a similar increase and
outperform bounded distance decoding $2^{r/2}/\sqrt{m\ln2}$ times.

\begin{corollary}
For $m\rightarrow\infty,$ soft decision recursive decoding of codes $\left\{
%TCIMACRO{\QATOP{\QTR{bf}{m}}{\QTR{bf}{r}}}%
%BeginExpansion
\genfrac{}{}{0pt}{}{\mathbf{m}}{\mathbf{r}}%
%EndExpansion
\right\}  $ corrects virtually all error patterns of Euclidean weight:
\begin{equation}%
\begin{array}
[c]{ll}%
\rho\leq\sqrt{n}(d/2m)^{1/2^{r}}, & \qquad\mbox{if }\;r=const,
\end{array}
\label{euc-low}%
\end{equation}%
\begin{equation}
\rho%
\begin{array}
[c]{ll}%
\leq\sqrt{n/m\ln2}, & \qquad\mbox{if }\;0<R<1.
\end{array}
\label{euc-high}%
\end{equation}

\end{corollary}

\section{Comparison}

In Table 1 we compare the asymptotic performance of the newly developed
algorithms with both the majority decoding and former recursive algorithms.
This comparison is done for hard and soft decision decoding of \ low-rate
codes of fixed order $r$ and for codes of fixed rate $R$. As above, for
\ low-rate codes we use the residual term $h$ of our error-correcting capacity
$n(1-h)/2$ for \ low-rate codes. Note that former recursive algorithms only
provided for bounded distance capacity $d/2$. For soft decision decoding of
low rate codes we use the squared Euclidean distance $\rho^{2}.$ Again, the
newly derived distance $\rho^{2}$ surpasses the one known for majority
decoding. Finally, for medium code rates $R$ we use the threshold weight $t,$
for which decoding yet corrects most error patterns. In this case, we double
decoding capacity of the former algorithms as seen from Table 1.

\medskip Table 1. Comparison of decoding algorithms for $\left\{
%TCIMACRO{\QATOP{\QTR{bf}{m}}{\QTR{bf}{r}}}%
%BeginExpansion
\genfrac{}{}{0pt}{}{\mathbf{m}}{\mathbf{r}}%
%EndExpansion
\right\}  $ codes\textbf{.}\nopagebreak

\bigskip%
\begin{tabular}
[c]{|c|c|c|c|}\hline%
\begin{tabular}
[c]{l}%
Decoding\\
capacity
\end{tabular}
&
\begin{tabular}
[c]{l}%
Former\\
Recursive
\end{tabular}
&
\begin{tabular}
[c]{l}%
Majority\\
Decoding
\end{tabular}
&
\begin{tabular}
[c]{l}%
New\\
Recursive
\end{tabular}
\\\hline%
\begin{tabular}
[c]{l}%
Hard decision,\\
$t$ for fixed $r$%
\end{tabular}
& $t=\frac{{\large d}}{{\large 2}}$ &
\begin{tabular}
[c]{l}%
$\frac{{\large n}}{{\large 2}}{\large \times(1-h),}$ \medskip\\
${\large h=(}\frac{{\large m}}{{\large n}}{\large )}^{{\large 1/2}%
^{{\large r+1}}}$%
\end{tabular}
&
\begin{tabular}
[c]{l}%
$\frac{{\large n}}{{\large 2}}{\large \times(1-h),}$ \medskip\\
${\large h=(}\frac{{\large m}}{{\large n}}{\large )}^{{\large 1/2}%
^{{\large r}}}$%
\end{tabular}
\\\hline%
\begin{tabular}
[c]{l}%
Soft decision,\\
$\rho^{2}$\ for fixed $r$%
\end{tabular}
& $\rho^{2}=\sqrt{{\large d}}$ & ${\large (}\frac{{\large n}}{{\large m}%
}{\large )}^{{\large 1/2}^{{\large r+1}}}\sqrt{{\large n}}{\Large \,}$ &
${\large (}\frac{{\large n}}{{\large m}}{\large )}^{{\large 1/2}^{{\large r}}%
}\sqrt{{\large n}}$\\\hline%
\begin{tabular}
[c]{l}%
Hard decision,\\
$t$\ for fixed $R$%
\end{tabular}
& $t=\frac{{\large d}}{{\large 2}}$ & ${\large d}\ln{\large d/4}$ &
${\large d}\ln{\large d/2}$\\\hline
\end{tabular}

\medskip Below in Table 2, we present simulation results for bit error rates
(BER) obtained by applying recursive and majority decoding to the code
$\left\{
%TCIMACRO{\QATOP{\QTR{bf}{9}}{\QTR{bf}{4}}}%
%BeginExpansion
\genfrac{}{}{0pt}{}{\mathbf{9}}{\mathbf{4}}%
%EndExpansion
\right\}  $ of length 512. We also exhibit the results of computer simulation
\ presented in \cite{bos}. Here, however, block error probabilities (BLER)
were used in recursive decoding.

Finally, the last row represents a refined version of recursive decoding. This
improvement uses the fact that recursive decoding gives different error rates
at different end nodes. In particular, the worst error rates are obtained on
the leftmost node $(m-r+1,1),$ and the next worst results are obtained at the
node $(m-r,1)$. This asymmetrical performance can be justified by our
qualitative analysis given above. We can see that the leftmost end node
operates at the highest noise power $\sigma^{2^{r}}.$

The next important conclusion is to set the corresponding information bits as
zeros. In this way we arrive at the subcodes of the original code $\left\{
%TCIMACRO{\QATOP{\QTR{bf}{m}}{\QTR{bf}{r}}}%
%BeginExpansion
\genfrac{}{}{0pt}{}{\mathbf{m}}{\mathbf{r}}%
%EndExpansion
\right\}  $ obtained by eliminating only a few least protected information
bits. This ``expurgation'' procedure gives a substantial improvement to
conventional recursive algorithms as seen from Table 2. Also, such a recursion
gives good block error probabilities (BLER) in contrast to most iterative
algorithms developed to date This part of the work performed jointly with K.
Shabunov has been developed further and will be reported separately in more detail.

\medskip Table 2. Decoding performance for code $\left\{
%TCIMACRO{\QATOP{\QTR{bf}{9}}{\QTR{bf}{4}}}%
%BeginExpansion
\genfrac{}{}{0pt}{}{\mathbf{9}}{\mathbf{4}}%
%EndExpansion
\right\}  .$\nopagebreak

\medskip%
\begin{tabular}
[c]{|l|l|l|l|}\hline
SNR (dB) & $2$ & $3$ & $4$\\\hline
Recursive \cite{bos} & $0.9$ & $0.5$ & $0.2$\\\hline
Majority \cite{dum3} & $0.3$ & $0.15$ & $0.1$\\\hline
Recursive (new) & $0.2$ & $0.03$ & $0.002$\\\hline
BER for subcodes & $0.05$ & $0.003$ & $3\cdot10^{-5}$\\\hline
BLER for subcodes & $0.2$ & $0.02$ & $2\cdot10^{-4}$\\\hline
\end{tabular}

\section{ Concluding remarks}

It is interesting to compare the presented recursive algorithm with a few
former variants considered in \cite{kab}, \cite{lit}, and \cite{bos}. Our
algorithm is similar to these especially to the one presented in \cite{kab}.
As a result, we achieve similar complexity of order $n\log n.$ However, there
are three differences. Instead of studying bounded-distance decoding, we try
to find actual decoding capacity of recursive algorithms. Secondly, in this
new setting we use probabilistic tools and explicitly recalculate posterior
probabilities while moving along the edges of Pascal triangle. Finally, we use
a different stopping rule and terminate any branch after reaching the codes
$\left\{
%TCIMACRO{\QATOP{\QTR{bf}{j}}{\QTR{bf}{1}}}%
%BeginExpansion
\genfrac{}{}{0pt}{}{\mathbf{j}}{\mathbf{1}}%
%EndExpansion
\right\}  $ of the first order$.$ The algorithms considered before were
terminated at codes $\left\{
%TCIMACRO{\QATOP{\QTR{bf}{j}}{\QTR{bf}{0}}}%
%BeginExpansion
\genfrac{}{}{0pt}{}{\mathbf{j}}{\mathbf{0}}%
%EndExpansion
\right\}  $ of order zero. By using probabilistic tools described above, one
can prove that using codes of order zero gives asymptotic performance similar
to that of majority decoding. Therefore our increase in decoding capacity
mostly results from a different stopping rule.

\end{document}